| | |
|---|---|
| Title | Effect of Cationic and Anionic Surfactants on the Application of Calcium Carbonate Nanoparticles in Paper Coating |
| Authors | Ahmed Barhoum[a,b] Hubert Rahier[a], Ragab Esmail Abou-Zaied[b,c], Mohamed Rehan[d], Thierry Dufour[e], Gavin Hill[a], and Alain Dufresne[b] |
| Affiliations | [a] Department of Materials and Chemistry, Vrije Universiteit Brussel, Pleinlaan 2, 1050 Brussels, Belgium<br>[b] The International School of Paper, Print Media and Biomaterials, Grenoble Institute of Technology (Grenoble INP)–Pagora, CS10065, 38402 Saint Martin d'Hères Cedex, France<br>[c] Cellulose and Paper Department, National Research Center (NRC), Dokki, Giza 11622, Egypt<br>[d] Fraunhofer Institute for Manufacturing Technology and Applied Materials Research (IFAM), Wiener Street 12, D-28359 Bremen, Germany<br>[e] Faculty of Sciences, Analytical and Interfacial Chemistry, Université Libre de Bruxelles, 2, Bvd du Triomphe, CP-255, B-1050 Bruxelles, Belgium |
| Ref. | ACS Appl. Mater. Interfaces, 2014, Vol. 6, Issue 4, 2734-2744 |
| DOI | http://dx.doi.org/10.1021/am405278j |
| Abstract | Modification of calcium carbonate particles with surfactant significantly improves the properties of the calcium carbonate coating on paper. In this study, unmodified and CTAB (hexadecyltetramethylammonium bromide)- and oleate-modified calcium carbonate nanoparticles were prepared using the wet carbonation technique for paper coating. CTAB (cationic surfactant) and sodium oleate (anionic surfactant) were used to modify the size, morphology, and surface properties of the precipitated nanoparticles. The obtained particles were characterized using X-ray diffraction (XRD), Fourier transform infrared (FT-IR) spectroscopy, zeta potential measurements, thermal gravimetric analysis (TGA), and transmission electron microscopy (TEM). Coating colors were formulated from the prepared unmodified and modified calcium carbonates and examined by creating a thin coating layer on reference paper. The effect of calcium carbonate particle size and surface modification on paper properties, such as coating thickness, coating weight, surface roughness, air permeability, brightness, whiteness, opacity, and hydrophobicity, were investigated and compared with commercial ground (GCC) calcium carbonate-coated papers. The results show that the obtained calcium carbonate nanoparticles are in the calcite phase. The morphology of the prepared calcium carbonate nanoparticles is rhombohedral, and the average particle diameter is less than 100 nm. Compared to commercial GCC, the use of unmodified and CTAB- and oleate-modified calcium carbonate nanoparticles in paper coating improves the properties of paper. The highest measured paper properties were observed for paper coated with oleate-modifed nanoparticles, where an increase in smoothness (decrease in paper roughness) (+23%), brightness (+1.3%), whiteness (+2.8%), and opacity (+2.3%) and a decrease in air permeability (−26%) was obtained with 25% less coat weight. The water contact angle at a drop age time of 10 min was about 112° for the paper coated with oleate-modified nanoparticles and 42° for paper coated with CTAB-modified nanoparticles compared to 104° for GCC-coated paper. |

# 1. Introduction

Paper is a highly versatile material with various favorable properties, for example, biodegradability, renewability, recyclability, mechanical flexibility, and affordability. Paper is mainly made of cellulosic pulp fibers derived from renewable natural bioresources including wood and non-wood lignocellulosic materials.1 The paper surface can remain somewhat rough and porous as it forms on the machine and begins to dry. Coating the paper surface with a color-containing pigment is an excellent method to impart certain qualities to the paper, including weight, surface smoothness, opacity, gas permeability, and reduced ink absorbency. The pigmented coating is very important when opacity, surface smoothness, or low gas permeability are needed at a low-basis weight. For food packing, the combination of low gas permeability and opacity is important to protect food from light and to prevent loss of the volatile contents. The color used for paper coating is a suspension of pigment and binder in water and has a solid content of 50–70%. This solid content is composed of 80–90 wt % pigment and 10–20 wt % binder.2 Pigment can fill in crevices and create a tight, flat, smooth surface that the addition of sizing or a perfect blend of fibers may not achieve. The binder attracts the pigment particles to each other and provides the required mechanical strength for the coat layer. Water allows the coat to be applied as a particulate suspension on the paper surface.

Different additives may be included in the color, such as thickener, dispersing agents, pH-controlling additives (i.e., acids or bases), lubricants, or biocides. A thickener is able to change both the rheology and water retention of the coating, and it can also affect the binding process. Dispersing agents and pH-controlling additives are added to ensure complete dispersion and long-term stabilization of the dispersed particles in the color.3 The basic pigments typically used in paper coating are ground and precipitated calcium carbonates, clays, silica, titanium dioxide, and talc. They typically are of micrometer size and consequently the thickness of the coatings is, even at its lowest, on the micrometer scale. Pigment shape, size, and size distribution are the main properties that determine the coat structure, its optical properties, and the resulting coated paper performance.4,5 Pigment particle shape is responsible for improving the coating structure through physical hindrance, whereas particle size and size distribution improve the coating structure through controlled consolidation. Particle size distribution and the shape of the pigments also determine the pore size and pore volume of the coating because of the variation in the packing characteristics of the pigment.5 Progress in nanotechnology has given way to the development of nanopigments for use in coatings, yet their exploitation has not been studied to a great extent. It is probable that the small particle size and high surface area confer a high surface quality (very smooth surface) and low gas permeability to the paper as well as adding new functions.

Currently, a number of studies have been devoted to special applications of nanopigments in papermaking.6 These are used in the preparation of high quality paper,7–13 low-gas-permeable paper,14 protected paper,15 hydrophobic paper,16–18 antimicrobial paper,19–24 photocatalytic paper,24–27 magnetic paper,28,29 electronic paper,30,31 and printed electronic paper.32,33 For example, Johnston et al. prepared nanostructured silica and silicate fillers for reducing print through and enhanced the print and optical properties of news print paper. Enomae and Tsujino produced spherical hollow nanoprecipitated calcium carbonate filler/coat for highly specific light-scattering paper.9 Nypelö et al. created an ultrathin and high-quality coat using color-containing precipitated calcium carbonate nanoparticles. 13 Kasmani et al. improved the surface smoothness and air-permeability characteristics of printing paper using a nanoclay/calcium carbonate-based coating color.14 Afsharpour et al. prepared a cellulosic-$TiO_2$ nanocomposite as a protective coating for old manuscript papers.15 Ogihara et al. prepared hydrophobic paper based on spray coating of hydrophobic silica nanoparticles on the paper surface.16 Ngo et al. found that the surface functionalization of paper with only a very small concentration of Au, Ag, and $TiO_2$ nanoparticles was able to produce devices with excellent photocatalytic and antibacterial properties.24 Small and Johnston used $Fe_2O_3$ nanoparticles as fillers to confer magnetic property to cellulosic paper.29 Anderson et al. reported the use of carbon nanotubes as fillers to produce electrically conductive paper.30 Ihalainen et al. created a thin narrow conductive coating ink from nano-Ag and organic polymer (polyaniline) for printing electronic applications. 32 Notwithstanding the importance of nanopigments, only a few grades of paper already contain nano calcium carbonate, and an industrial-scale breakthrough in this process has yet to come.34 The challenges in using nanopigments on the commercial scale are their poor dispersibility and the cost associated with their use.35 Nano calcium carbonate overcomes these challenges because of the availability of raw materials as well as the simplicity and low cost of its production.

Calcium carbonate can be produced by several routes such as wet carbonation,36 emulsion membranes,37 and high-gravity reactive precipitation.38 The wet carbonation route is an industrial route using calcium hydroxide and carbon dioxide gas for calcium carbonate production. Wet carbonation is a green chemistry approach because it is simple, inexpensive, consumes less material, avoids waste, and avoids the use of organic solvent.39

A wide variety of water-soluble additives has been successfully used for modifying calcium carbonate. Water soluble organic additives, cationic,[40,41] anionic,[42–52] and nonionic[53–55] or their mixtures,[56–59] were usually used during preparation to control the size, morphology, hydrophobicity, and dispersion stability of the produced calcium carbonate. Inorganic additives such as phosphoric acid,[60] sodium hexametaphosphate,[61] sodium silicate,[62] and fluorosilicic acid[63] can be used to improve the acid-tolerance properties of calcium carbonate. Other additives such as chitosan,[64] starch, and their derivatives[65] can be used to improve the bonding of calcium carbonate with polymers or fibers and consequently the strength properties of filled papers.

Surface modification of calcium carbonate particles can be performed either after synthesis, postmodification, or during the particle synthesis, in situ modification. The later can be realized more easily with higher efficiency than postmodification. In situ modification involves the addition of the modifier to the reaction system during or even prior to nanoparticle formation. In situ modification allows for the use of less modifier in addition to the advantages of controlling the particle size, morphology, and particle size distribution of the product. As already mentioned, calcium carbonate is currently available in different grades of anionic and cationic modification, which can be useful for different types of applications. However, there is still a lack in the fundamental understanding of the working principle. The insights gained in this article will help to optimize the coating. In this study, CTAB (cationic surfactant) and sodium oleate (anionic surfactant) were used during calcium carbonate preparation to prepare modified grades of calcium carbonate nanoparticles for high-quality and ultrathin paper coatings. CTAB or sodium oleate was added during the preparation to modify the size, morphology, and dispersion stability of calcium carbonate nanoparticles in coating color. The selection of CTAB and sodium oleate was based on the desire to compare a cationic surfactant to an anionic one as well as their white color, low cost, lack of toxicity, and similarities in molecular weight and structure. Previous work has shown that with addition of these substances nanoparticles could be obtained.[40,46] The effect of particle size, morphology, surface modification, and dispersibility of the produced particles in the water-based coating was investigated in terms of paper properties such as coating thickness and weight, surface roughness, air permeability, brightness, whiteness, opacity, contact angle, and wetting characteristics, which were compared to commercial ground calcium carbonate (GCC).

## 2. Experimental section

### 2.1. Materials

Analytical grade cetyltrimethyl ammonium bromide (CTAB) ($C_{19}H_{42}NBr$, 99+ %, Acros Organics), sodium oleate ($C_{18}H_{33}NaO_2$, 82+ % oleic acid, Sigma), calcium oxide (CaO, 97+ % on dry substance, Acros Organics), carbon dioxide gas ($CO_2$ gas, 99+ %, Air Liquide), and monodistilled water were used to prepare calcium carbonate particles. Paper with a basis weight of about 43.88 ± 0.50 g/m$^2$, thickness of about 59.8 ± 0.5 µm, area of 21.0 × 29.7 cm$^2$, and consisting of an 80% hardwood and 20% softwood pulp combination was used as a base for the coating application. Carboxylated styrene-butadiene latex (DL 966, >50% on dry substance, Dow Chemical Company), carboxy methyl cellulose sodium salt ($C_6H_9OCH_2COONa$, >99%, Fluka), tetra sodium diphosphate ($Na_4P_2O_7$, >95%, Sigma), and sodium hydroxide (NaOH, >97%, Sigma) were used to prepare the coating formulations. Commercial ground calcium carbonate (GCC) exhibiting calcite rhombohedral habit was used as the reference pigment for comparison.

## 2.2. Preparation of Calcium Carbonate

The wet carbonation method was used to prepare calcium carbonate nanoparticles. Four experiments were designed to prepare unmodified and CTAB- and oleate-modified calcium carbonates. The preparation conditions, CaO supersaturation concentration, CO2 flow rate, and surfactant concentration are shown in Table 1. The preparation of calcium carbonate was carried out in a plastic flask using the reaction system $Ca(OH)_2$–$H_2O$–$CO_2$, as described in the literature.66,67 The pure CaO reagent was reactivated via calcination at 1000 °C for 2 h. The calcinated CaO was slaked in monodistilled water containing the surfactant (CTAB or Oleate) at 80 °C. Then, the obtained lime was cooled to 25 °C. After cooling, pure $CO_2$ gas was blown into the lime milk from the bottom of the plastic bottle under vigorous stirring. The pH and electric conductivity of the reaction solution were inspected online using a pH meter (Jenway 3305) and electrical conductivity meter (Jenway 4510), respectively. When the pH value decreased to 9 and the electric conductivity showed a sharp decrease, the reaction was complete, and the $CO_2$ flow was stopped. The produced slurry was washed, filtered, and dried at 120 °C in an oven for 24 h to obtain calcium carbonate powder.

| sample | symbol | CaO (M)[a] | CO$_2$ flow rate (mL/min) | surface concentration (%)[a] |
|---|---|---|---|---|
| unmodified microsized CaCO$_3$ | MC | 1 | 100 | 0 |
| unmodified nano-CaCO$_3$ | NC | 1 | 1000 | 0 |
| CTAB-modified nano-CaCO$_3$ | CC | 1 | 100 | 2 |
| oleate-modified nano-CaCO$_3$ | OC | 1 | 100 | 2 |

*Table 1. Experimental Conditions for the Preparation of Calcium Carbonate. a: The CaO concentration was calculated as if the CaO is completely dissolved in the water, and the surfactant concentration is a weight percentage based on the expected CaCO$_3$ weight.*

## 2.3. Characterization of Calcium Carbonate

Phase identification of the prepared calcium carbonates samples was performed using an X-ray diffractometer (XRD, Bruker AXS D8, Germany) with Cu Kα (λ = 1.5406 Å) radiation and a secondary monochromator in the 2θ range from 20 to 70°. Bonding structures were analyzed using a Fourier transform infrared spectrometer (FTIR-460 plus, JASCO model 6100, Japan). KBr pellets (1:100 ratio) were used. Spectra were taken in the range of 4000–600 cm$^{-1}$ with a resolution of 4 cm$^{-1}$, and 64 scans were averaged per spectrum. The zeta potential of samples in suspension was measured at 25 °C using a Zeta meter 3.0 equipped with a microprocessor unit (Malvern Instrument Zetasizer 2000). The unit automatically calculates the electrophoretic mobility of the particle and converts it into zeta potential using the Smoluchowski equation.68 Thermogravimetric analysis was performed on a TGA Q5000. The samples were dried isothermally at 55 °C for 20 min before heating from 55 to 1000 °C at a heating rate of 10 °C/min under an air atmosphere (50 mL/min). High-temperature platinum pans were used, and the sample mass was approximately 7 mg. The morphology of the calcium carbonate samples and ground calcium carbonate was investigated using a high-resolution analytical transmission electron microscope (TEM, Jeol JEM-2010, Japan) operating at a maximum of 200 kV.

## 2.4. Preparation of Coating Colors

Calcium carbonate pigments, ground calcium carbonate (GCC), and the four prepared calcium carbonates, MC, NC, OC, and CC, were used to prepare five colors. The colors were formulated at 100 parts of pigment, 15 pph (parts per hundred) of styrene butadiene (SB) latex as binder, and 8 pph of carboxymethylcellulose sodium salt (CMC) as thickener. Additionally, 3 pph of tetra sodium diphosphate salt was used as dispersant. The required weights of calcium carbonate, tetra sodium diphosphate salt, and water were put in a beaker and agitated. CMC powder was added slowly to the pigment slurry to avoid any viscosity shock or lump formation. The slurry was agitated at high speed for complete dispersion of CMC. The speed of the agitator was slowed to avoid any foam formation during addition of synthetic binder. The pH of the color was adjusted to 9 using a 5 M NaOH solution. The total solid content of the colors was kept around 50%. It should be emphasized here that there are difficulties in redispersing the unmodified calcium carbonate powders into the coat formulation. This is because the particles strongly aggregate when they are dried. For this reason, all pigments are first ground by mortar and pestle and then ultrasonically dispersed to redisperse the pigments during the coating color preparation. All coating colors were dispersed under ultrasonication for 30 min to ensure complete nano calcium carbonate dispersion.

## 2.5. Application of Coating Colors

Reference papers were preconditioned for 24 h at 25°C and 65% relative humidity. The prepared coating colors were applied on the reference papers using a universal coating machine (see Supporting Information Figure 1s) developed by CTP (Centre Technique du papier, Grenoble, France). The main operative conditions were as follows: velocity, 3 m/min; IR drying; drying time, 1 min; and roll coater number zero graduation. Reference papers were coated on one side at a controlled pressure and fixed coating speed of 3 m/min. After the limit of its travel, the coated papers were dried in an IR-drying chamber for 1 min.

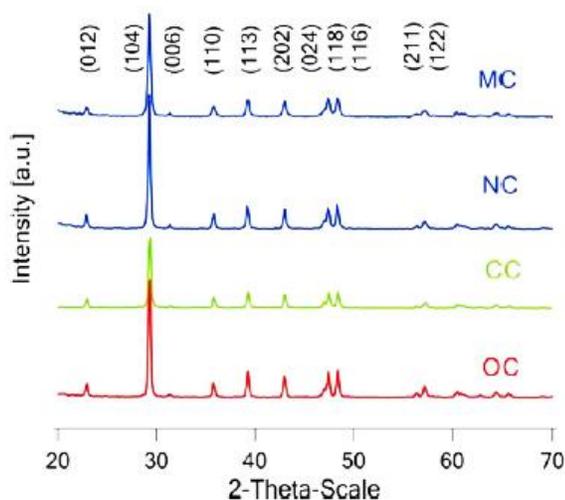

*Figure 1. X-ray diffractions patterns of the prepared calcium carbonates: MC, NC, CC, and OC.*

## 2.6. Characterization of Coated Papers

The physical and optical properties of the reference and coated papers were determined by conditioning the paper sheets at 25°C and 65% relative humidity for more than 24 h. After that, all paper properties were measured using standard methods of testing. Standard deviation for the measurements was calculated on the basis of 10 replicates for each sample.

**2.6.1. Scanning Electron Microscopy (SEM)**

Scanning electron microscopy (SEM, Jeol-JSM-5410, Japan) was used to study the surface morphology of the GCC reference filler as well as the reference and coated papers. The test samples were mounted on specimen stubs with double-sided adhesive tape, coated with gold/palladium in a sputter coater, and examined by SEM at an accelerating voltage of 10 kV with a tilt angle of 45°.

**2.6.2. Paper Thickness and Grammage**

The thickness was measured from 10 paper sheets on the basis of Tappi 411 using a L&W micrometer 51 with an accuracy of 0.1 µm. Then, the grammage (basis weight) was determined according to the Tappi standard (Tappi 410). The weight of each tested sample was taken separately on a precision balance with an accuracy of 0.1 mg, and grammage was calculated as follows: G =1000*M/A where G = grammage (g/m$^2$), M = weight of the sheet (g), and A =area of the sheet (cm$^2$).

**2.6.3. Coating Thickness and Weight**

The thickness of the coating (µm) on the paper surface was calculated by subtracting the thickness of a coated area of coated paper from the thickness of the same area of the uncoated paper. The coating thickness was calculated according to the following equation.

$$T = T_2 - T_1 \qquad (2)$$

where T = coating thickness (µm), $T_2$ = thickness of coated paper (µm), and $T_1$ = thickness of reference paper (µm).

The coating weight (g/m$^2$) was determined gravimetrically from the weight difference between a coated and uncoated paper sample having an area of 10 × 10 cm$^2$ (balance, 0.1 mg).69 The coating weight was calculated according to the following equation

$$W = G_2 - G_1 \qquad (3)$$

where W = coating weight (g/m$^2$), $G_2$ = grammage of coated paper (g/m$^2$), and $G_1$ = grammage of reference paper (g/m$^2$).

**2.6.4. Coat Ash Content**

Thermogravimetric analysis was performed to determine the calcium carbonate content in the coated paper as a second method for checking the coating thickness and weight. The calcium carbonate content in the coated papers was calculated from the coat ash content percentage for coated papers at 600 °C, which was chosen because it is the temperature at which the reference paper and organic part of the coat completely decomposed while the decarbonation of the carbonate did not start yet. The samples were dried isothermally at 60 °C for 5 min before heating from 60 to 700 °C at a heating rate of 10 °C/min under an air atmosphere (50 mL/min). Platinum pans were used, and sample mass was approximately 9 mg. The coat ash content percentage for the coated paper was calculated according to the following equation

$$A = A_2 - A_1 \qquad (4)$$

where A = coat ash content (%), $A_2$ = ash content of the coated paper (%), and $A_1$ = ash content of the reference paper (%).

**2.6.5. Surface Roughness and Air Permeability Measurements**

The paper surface roughness and air permeability values of the paper sheets were measured using a Bendtsen ME-113 roughness and air permeance tester based on ISO 5636:3. Surface roughness has an important influence on the printing quality. Roughness also affects properties such as the coefficient of friction, gloss, and coating absorption. The surface roughness for the coated paper was determined by measuring the air flow between the sample paper (backed by flat glass on the bottom side) and two pressurized, concentric annular lands that are impressed into the sample from the top side. The rate of air flow (mL/min) is related to the surface roughness or smoothness of paper. Air permeability was determined by measuring the rate of air flow under standard pressure between the paper surface and two concentric, annular metal rings applied to the paper.70 The air permeability of a paper web is a physical parameter that characterizes the degree of web resistance to air flow. The air permeability of coated papers is partially dependent on the uniformity and porosity of the coating layer. Air permeability is a critical property of papers used in food packaging. Water vapor and $O_2$ are critical compounds that can penetrate through the packaging materials and degrade food quality.

**2.6.6. Optical Properties Measurements**

The optical properties of paper, brightness, whiteness, and opacity, were evaluated using standard testing for physical and optical properties. All optical tests were qualitative and conducted using an integrating refractometer, model JY9800. The instrument is multifunctional; it can be used to measure brightness, whiteness, and opacity. The brightness, whiteness, and opacity of reference and coated papers were measured according to ISO 2470:1999, ISO 11475:1999, and ISO 2471:1998, respectively. Whiteness differs fundamentally from paper brightness. Brightness may or may not add much value to the "useful" properties of the paper. Brightness is defined as the percentage reflectance of blue light only at a wavelength of 457 nm. Whiteness refers to the extent that paper diffusely reflects light for all wavelengths throughout the visible spectrum. So, whiteness is an appearance term. Opacity is the measure of how much light is kept away from passing through a sheet. A perfectly opaque paper is the one that is absolutely impervious to the passage of all visible light. It is the ratio of diffused reflectance and the reflectance of a single sheet backed by a black body. The opacity of paper is influenced by numerous factors such as thickness, amount and kind of filler, and coating pigments.71

**2.6.7. Contact Angle Measurements**

The water contact angle (WCA) of the reference and coated paper was measured with a Kruss DSA-100 contact angle analyzer. The contact angle of water on the substrate was calculated on the basis of a numerical solution of the full Young–Laplace equation by a computer program from the equipment supplier. All contact angle measurements were carried out at 25 °C every minute from the profile of the droplets that were fully separated from the pump syringe needle tip. The droplet volume was 5 μL, and at least three parallel measurements were recorded.

# 3. Results & Discussion

## 3.1. Effect of Surfactant Concentration and $CO_2$ Flow Rate on the Calcium Carbonate

The effect of surfactants concentration and $CO_2$ flow rate and on the polymorphism, particle size, morphology, and particle surface charge of the prepared calcium carbonates was investigated. X-ray diffraction patterns of the prepared calcium carbonate samples, MC, NC, CC, and OC, are shown in Figure 1. The diffraction peaks at the characteristic 2θ positions can be indexed to the different planes of calcite crystals.72 Figure 2 shows the typical FT-IR spectra of the calcium carbonate samples, MC, NC, CC, and OC. All prepared samples show characteristic absorption peaks of $CO_3^{2-}$ appearing at 3445, 2515, 2355, 1447, 877, and 712 cm$^{-1}$. The broad absorption peaks at 3445 are assigned to the stretching vibration and asymmetric stretching vibration of the O–H bond and can be attributed to the presence of absorbed water and hydroxyl groups on the surface of calcium carbonate particles. The peaks around 2355 cm$^{-1}$ are attributed to carbon dioxide in the atmosphere. The peaks at 2515, 1447, 877, and 712 cm$^{-1}$ are ascribed to calcium carbonate. The strong (saturated) absorption peak at 1447 cm$^{-1}$ is assigned to the asymmetric stretching vibration of the C–O bond. The absorption peaks at 877 and 712 cm$^{-1}$ are assigned to the bending vibration of the C–O bond. The combination of the three peaks at 1447, 877, and 712 cm$^{-1}$ appears at 2515 cm$^{-1}$. According to the IR standard spectrum of calcium carbonate,73 the prepared calcium carbonate samples, MC, NC, CC, and OC, are typically calcite crystals, which is consistent with the XRD analysis results. By comparing the MC, NC, CC, and OC samples in Figure 2 (inset), a shoulder at 1615 cm$^{-1}$ is observed for OC. It corresponds to the appearance of a carboxylic salt, indicating that oleate has been attached to the surface of calcium carbonate via an ionic bond. The peaks at 2955, 2925, and 2885 cm$^{-1}$ are ascribed to the long alkyl chain of oleate (CH stretching region) and prove the presence of oleate at the surface of calcium carbonate.74 In the case of the CC sample, peaks at 2955, 2923, and 2846 cm$^{-1}$ ascribed to the long alkyl chain of CTAB further prove the presence of CTAB at the surface of calcium carbonate. The FT-IR spectra of pure CTAB and sodium oleate are given in Figure 2s (Supporting Information). The XRD and FT-IR results indicate that no characteristic absorption of other phases is observed, indicating that CTAB and sodium oleate have no significant effect on calcium carbonate polymorphism.

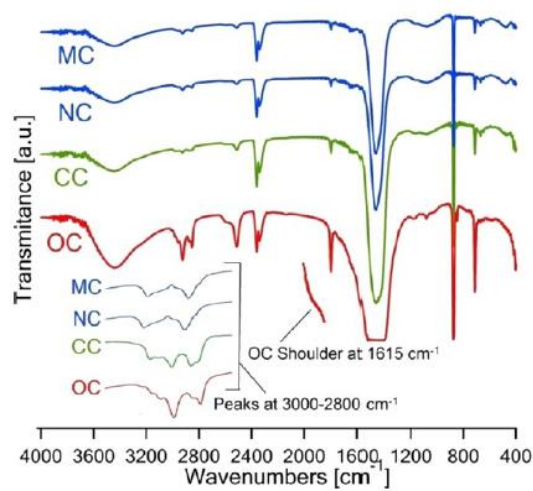

*Figure 2. FT-IR spectra of the prepared calcium carbonates: MC, NC, CC, and OC.*

Zeta potentials of MC, NC, CC, and OC were determined at pH 9. The zeta potentials of the prepared samples are −14.5, −15.6, −0.5, and −22 mV, respectively. The results indicate that the surface charge of calcium carbonate decreases to more negative values with addition of 2 wt % sodium oleate and increases to less negative values with addition of 2 wt % CTAB compared to unmodified samples. The change in surface potential of calcium carbonate is due to adsorption of CTAB and sodium oleate on the surface of the precipitated particles. TGA measurements of the CTAB- and oleate-modified calcium carbonate nanoparticles confirm that both CTAB and oleate are successfully adsorbed on the calcium carbonate surface. However, the amount of adsorbed oleate was much higher than CTAB. The mass-loss measurements revealed that the amount of adsorbed CTAB and oleate reaches up to 20 and 75% compared to the added amount of the surfactants, respectively. This could be explained by the ability of the carboxylic group (−COO−) of oleate to bind effectively to the $Ca^{2+}$ ions on the calcium carbonate particle surface with ionic bonds. However, CTAB, which is a quaternary ammonium compound ($NR_4^+Br^-$), can only interact with the calcium carbonate particle via van der Waals forces. Consequently, CTAB is less efficiently adsorbed on calcium carbonate particles. Adsorption of CTAB neutralizes the negative surface charge on calcium carbonate and increases the surface charge to values close to the isoelectric point (IEP) of calcium carbonate, whereas adsorption of the oleate anion on calcium carbonate particles imparts a more negative surface charge to calcium carbonate.

Figure 3 shows the effect of the preparation conditions, $CO_2$ flow rate, and surfactant concentration on the particle size and morphology of calcium carbonate. It can be seen in Figure 3 that the use of a CaO concentration of 1 M and low $CO_2$ flow rate of 100 mL/min forms scalenohedral particles of 300–500 nm in diameter and 1.5–2 μm in length. These particles are thus micro size calcium carbonate, MC. Upon increasing the CO2 flow rate to 1000 mL/min at the same CaO concentration, the calcium carbonate particle size is reduced to 60–100 nm. Under these conditions, it is thus possible to form rhombohedral nanoparticles, NC. This is explained by the acceleration of the carbonation reaction leading to nanoparticles. The excess of $Ca^{2+}$ in the bulk solution and a low $CO_2$ flow rate elongate the reaction and growth time of precipitated particles, leading to a larger particle size.75 Because of the addition of CTAB and oleate (up to 2 wt %), a lower flow rate of 100 mL/min $CO_2$ can again be used while still obtaining rhombohedral nanoparticles. Under the given conditions of 1 M CaO, the particle size is decreased to 20–35 nm in width and 40–80 nm in length. CTAB and oleate can be situated at the gas–liquid interface. Consequently, they increase the stability of $CO_2$ bubbles and prevent their aggregation. The maintenance of $CO_2$ bubbles enhances the mass transfer of $CO_2$ into solution and increases the $CO_3^{2-}/Ca^{2+}$ ionic ratio. Moreover, adsorption of CTAB or oleate on the calcium carbonate particle surface can cause inhibition of crystal growth and decrease the particle size. By comparing commercial GCC with the prepared calcium carbonate, the polymorphism, zeta potential, particle size, and morphology of GCC were investigated. The results show that GCC has a calcite polymorphism, particle surface charge of about −14.5 mV at pH 9, rhombohedral particle morphology, and particle size from 0.01 to 2.5 μm. The particle size and morphology of GCC was investigated using TEM and SEM, as shown in Figure 4.

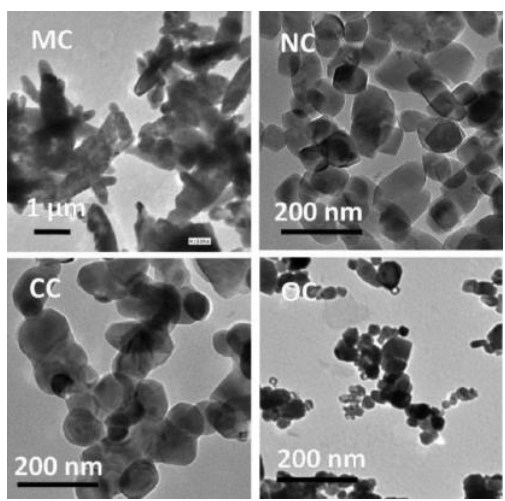

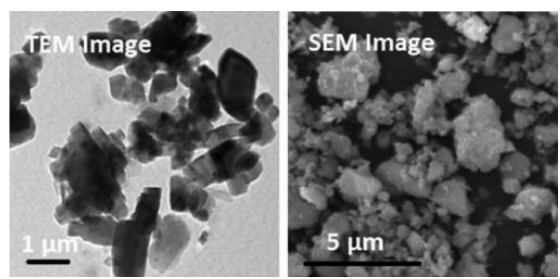

Figure 4. TEM and SEM images of the commercial ground calcium carbonate GCC.

Figure 3. TEM images of the prepared calcium carbonates: MC, NC, CC, and OC.

## 3.2. Properties of Coated Papers

The effect of nanosize calcium carbonate and modification with CTAB and oleate on the physical and optical properties were investigated by creating a thin coating layer from the prepared calcium carbonate colors on the paper surface. The obtained papers coated with MC, NC, CC, and OC were evaluated by standard physical and optical tests and compared with commercial GCC (Tables 2 and 3).

**3.2.1. Coating Morphology**

The surface microstructure of reference and coated papers was studied to understand how the calcium carbonate nanoparticle coating can reduce paper surface irregularities and improve paper properties. SEM images in Figure 5 show that the nanoparticles of calcium carbonates, NC, CC, and OC, can fill the pores on the reference paper surface and reduce surface irregularities more efficiently than the GCC and MC particles. It is striking that the number and size of cracks on the surface is much smaller when using nanoparticles than when using MC or GCC. The prepared calcium carbonate nanoparticles have a rhombohedral morphology and size distribution in the range of 20–100 nm (Figure 3). MC and GCC particles exhibit scalenohedral and rhombohedral morphology, respectively, with a broader size distribution, as shown in Figures 3 and 4. The difference in particle size and morphology between the nano calcium carbonate (NC, OC, and CC) and the microsize ones (MC and GCC) significantly affects the coating structure and the way particles pack together on the paper surface. Calcium carbonate nanoparticles can pack significantly better with minimum interparticle voids compared with the microsize particles when the coat porosity and pore size are correlated to the particle size and size distribution of the coating pigments.[76,77] During coating, the rhombohedral morphology of the prepared nanoparticles can reduce the friction between the coating rod and paper surface and produce a smoother surface than with GCC and MC. Rhombohedral nanoparticles can roll over each other and decrease the friction between the coating rod and the paper surface. This could explain the high surface smoothness of the paper coated with NC, CC, and OC. By comparing NC and CC with OC, it is clear that the surface of OC is smoother than that of NC and OC. This could be due to the difficulties to redispersing the NC particles into the coating formulation. For CC, the cationic modification with CTAB may decrease the dispersion stability of the CC particles in the coating formulation and cause aggregation. The CC nanoparticles together with the anionic additives of the coating color, such as carboxymethylcellulose sodium salt (CMC) and tetra sodium diphosphate, tend to form aggregates and decrease the coating color stability, whereas the anionic modification with oleate improves the dispersibility of the calcium carbonate nanoparticles and the stability of the coating color. For OC, adsorption of oleate on the particle surface can decrease the particles surface potential to a more negative potential and increase the dispersibility, stability, and efficiency of calcium carbonate in the coating color. The high stability and good dispersion of oleate-modified calcium carbonate particles effectively improve the paper surface properties compared to unmodified and CTAB-modified nano calcium carbonate (Tables 2 and 3).

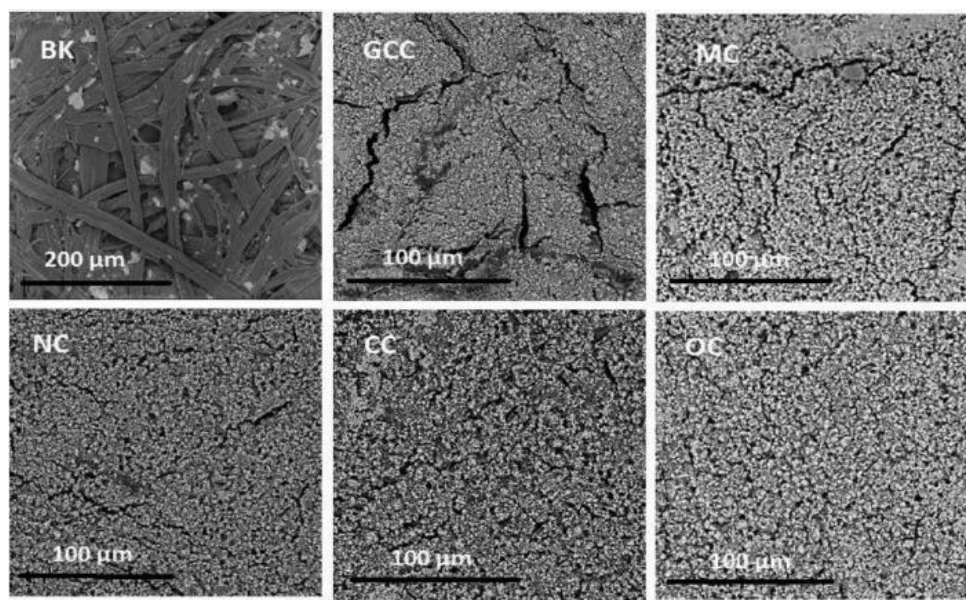

*Figure 5. SEM images of the reference paper (BK) and coated papers (GCC, MC, NC, CC, and OC).*

### 3.2.2. Coating Thickness and Weight

The variation in pigment size, shape, and modifications can significantly affect the rheological behavior of coating color as well as the coating thickness and weight. On the basis of the average grammage of reference paper, the coating thickness and weight of papers coated with commercial GCC and the prepared nanosized calcium carbonates were calculated and are represented in Table 2. The results show that the nano calcium carbonates, NC, CC, and OC, coating thickness and weight are approximately 6−20 and 15−26% lower than those with GCC, respectively. However, the coating thickness and weight for MC-coated papers are approximately 16 and 5% higher than with GCC, respectively. This result clearly shows that with a smaller particle size and rhombohedral morphology of calcium carbonate a substantially thinner coat layer and lower coating weight can be obtained even under the same coating conditions. The change in the particle size and morphology of calcium carbonate from microsize scalenohedral morphology to nanosize causes the observed decrease in the coat thickness and weight from MC to NC, CC, and OC.[4,9,77]

| samples | symbols | paper physical properties ± SD | | | |
|---|---|---|---|---|---|
| | | coating thickness ($\mu$m) | coating weight (g/m$^2$) | paper roughness (mL/min) | air permeability (mL/min) |
| reference paper | BK | | | 584 ± 21 | 955 ± 23 |
| fine GCC | GCC | 5.1 ± 0.8 | 6.74 ± 0.98 | 505 ± 35 | 592 ± 25 |
| unmodified micro-CaCO$_3$ | MC | 5.9 ± 0.8 | 7.10 ± 0.96 | 458 ± 24 | 550 ± 20 |
| unmodified nano-CaCO$_3$ | NC | 4.8 ± 0.6 | 6.10 ± 0.99 | 397 ± 19 | 443 ± 14 |
| CTAB-modified nano-CaCO$_3$ | CC | 4.4 ± 0.7 | 5.30 ± 0.84 | 460 ± 15 | 537 ± 15 |
| oleate-modified nano-CaCO$_3$ | OC | 4.1 ± 0.6 | 5.02 ± 0.89 | 388 ± 16 | 439 ± 18 |

Table 2. Physical Properties of Reference Paper, GCC-Coated Paper, and Paper Coated with the Prepared Calcium Carbonate

| samples | symbols | paper optical properties ± SD | | |
|---|---|---|---|---|
| | | brightness (%) | whiteness (%) | opacity (%) |
| reference paper | BK | 86.25 ± 0.71 | 84.19 ± 0.71 | 70.45 ± 0.71 |
| ground calcium carbonate | GCC | 87.80 ± 0.73 | 85.94 ± 0.73 | 77.21 ± 0.65 |
| unmodified micro-CaCO$_3$ | MC | 88.10 ± 0.60 | 87.45 ± 0.65 | 78.70 ± 0.55 |
| unmodified nano-CaCO$_3$ | NC | 88.90 ± 0.66 | 87.71 ± 0.66 | 77.90 ± 0.56 |
| CTAB-modified nano-CaCO$_3$ | CC | 88.64 ± 0.64 | 88.35 ± 0.94 | 77.97 ± 0.45 |
| oleate-modified nano-CaCO$_3$ | OC | 88.35 ± 0.51 | 87.95 ± 0.66 | 78.95 ± 0.56 |

Table 3. Optical Properties of Reference Paper, GCC-Coated Paper, and Paper Coated with the Prepared Calcium Carbonates

### 3.2.3. Air Roughness and Permeability

The Bendtsen air roughness and permeability of reference paper (BK), GCC coated paper, and paper coated with the prepared calcium carbonates, MC, NC, CC, and OC, are listed in Table 2. The obtained results show that Bendtsen roughness and air permeability of papers coated with nano calcium carbonates NC, CC, and OC are 9−23 and 9−26% lower than for GCCcoated paper, respectively. The Bendtsen roughness and air permeability of papers coated with MC particles are about 9 and 7% lower than those for GCC-coated paper, respectively. These results indicate that the nanosize and rhombohedral morphology of calcium carbonate samples (NC, CC, and OC) lead to a compact coating structure filling the crevices and create a tight, flat, smooth surface with less air permeability than those of microsize particles GCC and MC.

### 3.2.4. Brightness, Whiteness, and Opacity

The brightness, whiteness, and opacity of reference paper (BK), paper coated with the commercial GCC, and paper coated with the prepared calcium carbonates, MC, NC, CC, and OC, are summarized in Table 3. Papers coated with nano calcium carbonates NC, CC, and OC show an improvement in brightness (0.6–1.3%), whiteness (2.1–2.8%), and opacity (0.9–2.3 %) compared to commercial GCC. This increase is substantial when taking into account the coat weight of nanosized calcium carbonates, which is about 15–26% less than the one of commercial GCC. The calcium carbonate contents in the coated paper were determined with TGA (Figure 6). The results show that the calcium carbonate contents in the GCC-, MC-, NC-, CC-, and OC-coated papers are about 10.1, 12.40, 7.9, 7.1, and 6.7%, respectively. These results could explain the small increase in the optical properties of the nano-coated paper compared with GCC and MC. The small particle size of the prepared nano calcium carbonate samples results in a change in the geometry of the pores, which affects the optical properties of the coated paper and increases the paper brightness, whiteness, and opacity compared with GCC-coated paper. The high opacity of micro calcium carbonate MC compared with the prepared nano calcium carbonates and GCC may be attributed to the scalenohedral particle morphology and high coat thickness of MC. The scalenohedral morphology of MC may provide optimum internal particle voids in particles filled with air, which probably leads to the high light-scattering ability, contributing to opacity. These results are consistent with the literature.[4,77] The improvement in optical properties upon using nanoparticles is quite significant, but further improvements can still be expected via optimization of the coating color composition and control of the pigment content, coating thickness, and weight.

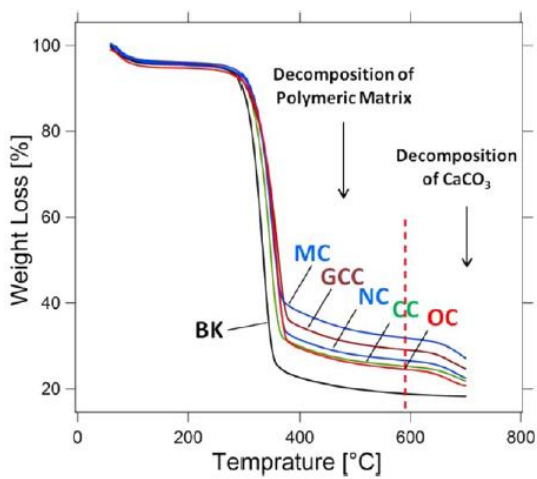

Figure 6. TGA of the reference paper (BK) and coated papers (GCC, MC, NC, CC, and OC).

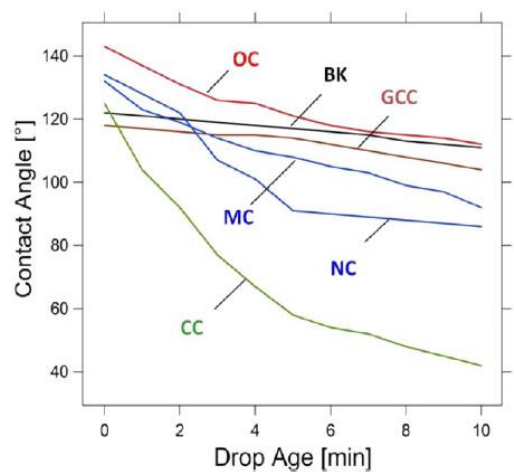

Figure 7. Time dependency of the contact angle of a water droplet on the reference paper (BK) and coated papers (GCC, MC, NC, CC, and OC).

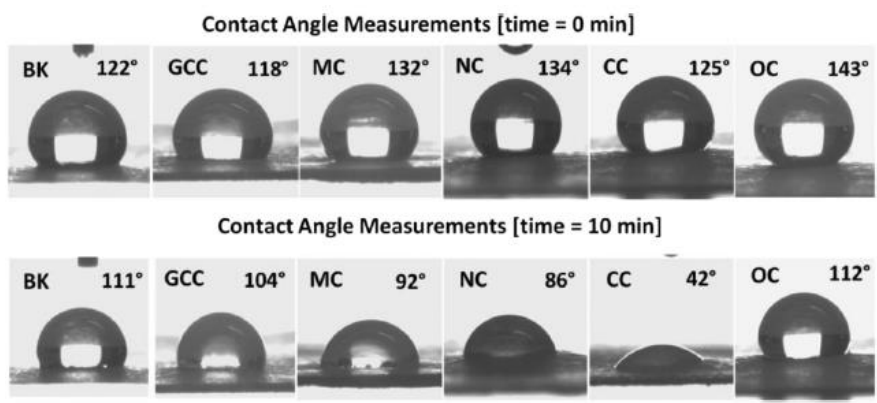

Figure 8. Water contact angles of the reference paper (BK) and coated papers (GCC, MC, NC, CC, and OC) at 0 and 10 min.

**3.2.5. Water Droplet Behavior on the Coat Surface**

The wetting behavior and surface hydrophobicity of the reference paper BK and the papers coated with GCC, MC, NC, CC, and OC were characterized by contact angle measurements, as shown in Figure 7 and 8. The values of the contact angle versus time are listed in Table 2s (Supporting Information). The values of contact angles given by Figure 7 and Table 1s show that the hydrophobicity of the coated paper increases in the order of CC < NC < MC < GCC < OC. The results indicate that the hydrophobicity of the coated paper significantly change with the particle size, morphology, and surface properties of the coating pigments and the obtained coating structure. The use of unmodified nanoparticles NC decrease the paper hydrophobicity compared with the microsize particles GCC and MC. The surface modification with oleate acts against the particle size effect and increases the hydrophobicity to the ultrahydrophobic level. The decrease in the hydrophobicity for CC paper compared to NC and OC paper is due to the coating structure of the CC. The coating structure of CC paper appears more open with a bigger pore size compared to NC and OC, and the instability of the CC coating color produces a more open coating structure and allows more water to be absorbed compared to NC and OC.

In paper printing applications, the absorbtion of the inks often leads to a numbers of print quality problems. For uncoated, unsized paper, the inks are in contact with wood fibers, which promotes fluids penetration into the sheet. This penetration leads to a low print density and high print through. Papers intended for inkjet printing (water-based inks) are often sized to reduce ink penetration, but too much sizing may also lead to a decrease in the printing density. Therefore, it may be beneficial to investigate a new material to control the penetration of inks into uncoated paper. For this purpose, we suppose that the paper coated with OC will reduce the tendency of the water-based inks to be spread over the paper surface, this could improve the brilliant appearance of the print. However, the coating with CC can also be used for paper grades that have a low affinity towards water-based inks. The paper coated with CC could enable the anionic water-based inks to fix on the paper surface and dry fast.78 However, it would be ideal for the paper to have a double layer of surface coating from an OC inner coating layer and a CC outer layer. The outer layer CC can fix the inks on the paper surface, whereas the interior layer OC can prevent the ink penetration and print through. Controlling the thickness of outer and interior layer the printing properties of light-weight papers could be improved.

# 4. Conclusions

CTAB (cationic surfactant) and sodium oleate (anionic surfactant) an play important role in the preparation and application of calcium carbonate nanoparticles in paper coating. Addition of CTAB or oleate (up to 2 wt %) during preparation inhibits particle growth and leads to rhombohedral particles of 20–100 nm rather than microsized scalenohedral particles. Oleate bonds strongly on the prepared nanoparticle surface through ionic bonds, whereas CTAB interacts with the particles only via van der Waals forces. Consequently, oleate is more effective than CTAB at modifying the size, morphology, and surface properties of the calcium carbonate particle. Redispersing nano calcium carbonate in water is problematic, but anionic modification with oleate affects the particle surface potential and improves the dispersibility of the particles in the coating colors. This improves the efficiency of calcium carbonate coating on the paper surface. Cationic modification with CTAB decreases the coating color stability and consequently it decreases the efficiency of calcium carbonate coating on paper surface. Oleate-modified calcium carbonate can significantly improve paper smoothness (+23%), brightness (+1.3%), whiteness (+2.8%), and opacity (+2.3%) and decrease air permeability (−26%) compared to commercial GCC even if a decrease in the coating weight of 25% is observed. Coating the paper surface with oleate-modified nano calcium carbonate can significantly change the paper surface from hydrophilic to hydrophobic. The highest measured water contact angle for the nanoparticle-coated paper was observed for oleate-modified nanoparticles, and the lowest contact angle was observed for CTAB-modified nanoparticles. The contact angle at a drop age time of 10 min was about 112° for the paper coated with oleatemodified nanoparticles and 42° for paper coated with CTAB modified nanoparticles compared to 104° for GCC-coated paper. Future work is needed to fine tune the coatings for optimal printability. For instance, the results of this work suggest that the use of a double layer, namely, a bottom oleate-modified nano carbonate layer for opacity and a top CTAB-modified nano carbonate for printability, might substantially improve paper quality.

# 5. Acknowledgements

The present study is a part of Ahmed Barhoum's Ph.D. work, which was partially financially supported by the French Culture Center in Cairo (grant no. 759302C) and the Medastar Erasmus Mundus Program (grant no. 2011-4051/002-001-EMA2). A.B. thanks Associate Prof. Said Elsheikh (Central Metallurgical Research and Development Institute, Egypt), Associate Prof. Samya El-Sherbiny and Associate Prof. Fatma Morsy (Printing and Packaging Lab., Helwan University, Egypt), Prof. Waleed El-Zawawy (Cellulose and Paper Department, National Research Center, Egypt), and Prof. Naceur Belgacem (The International School of Paper, Print Media and Biomaterials, France) for their help and valuable discussions.

# 6. Abbreviations

CTAB, hexadecyltetramethylammonium bromide

XRD, X-ray diffraction

FT-IR, Fourier transform infra-red

GCC, ground calcium carbonate

TGA, thermogravimetric analysis

TEM, transmission electron microscopy

SB, styrene butadiene

CMC, carboxymethylcellulose sodium salt

# 7. References

(1) Hubbe, M. A.; Bowden, C. Handmade Paper: A Review of Its History, Craft, and Science. BioResources 2009, 4, 1736–1792.

(2) Laudone, G. M.; Matthews, G. P.; Gane, P. A. Modelling the Shrinkage in Pigmented Coatings during Drying: A Stick–Slip Mechanism. J. Colloid Interface Sci. 2006, 304, 180–90.

(3) Smook, G. A. Handbook for Pulp & Paper Technologists, 3rd ed.; Angus Wilde Publications: Vancouver, Canada, 2002; pp 288–295.

(4) Kumar, N.; Bhardwaj, N. K.; Chakrabarti, S. K. Influence of Pigment Blends of Different Shapes and Size Distributions on Coated Paper Properties. J. Coat. Technol. Res. 2011, 8, 605–611.

(5) Kumar, N.; Bhardwaj, N. K.; Chakrabarti, S. K. Influence of Pigment Blends of Different Shapes and Size Distributions on Coated Paper Properties. J. Coat. Technol. Res. 2011, 8, 613–618.

(6) Shen, J.; Song, Z. Q.; Qian, X. R.; Ni, Y. H. A Review on Use of Fillers in Cellulosic Paper for Functional Applications. Ind. Eng. Chem. Res. 2011, 50, 661–666.

(7) Park, J. K.; Kim, J. K.; Kim, H. K. TiO2–SiO2 Composite Filler for Thin Paper. J. Mater. Process. Technol. 2007, 186, 367–369.

(8) Johnston, J. H.; McFarlane, A. J.; Borrmann, T.; Moraes, J. Nano-Structured Silicas and Silicates–New Materials and Their Applications in Paper. Curr. Appl. Phys. 2004, 4, 411–414.

(9) Enomae, T.; Tsujino, K. Application of Spherical Hollow Calcium Carbonate Particles as Filler and Coating Pigment. Appita J. 2004, 57, 493–493.

(10) Lattaud, K.; Vilminot, S.; Hirlimann, C.; Parant, H.; Schoelkopf, J.; Gane, P. Index of Refraction Enhancement of Calcite Particles Coated with Zinc Carbonate. Solid State Sci. 2006, 8, 1222–1228.

(11) Koivunen, K.; Niskanen, I.; Peiponen, K. E.; Paulapuro, H. Novel Nanostructured PCC Fillers. J. Mater. Sci. 2009, 44, 477–482.

(12) Koivunen, K.; Paulapuro, H. Papermaking Potential of Novel Nanostructured PCC Fillers. Appita J. 2010, 63, 258–258.

(13) Nypelo, T.; Osterberg, M.; Laine, J. Tailoring Surface Properties of Paper Using Nanosized Precipitated Calcium Carbonate Particles. ACS Appl. Mater. Interfaces 2011, 3, 3725–3731.

(14) Kasmani, J. E.; Mahdavi, S.; Alizadeh, A.; Nemati, M.; Samariha, A. Physical Properties and Printability Characteristics of Mechanical Printing Paper with LWC. BioResources 2013, 8, 3646–3656.

(15) Afsharpour, M.; Rad, F. T.; Malekian, H. New Cellulosic Titanium Dioxide Nanocomposite as a Protective Coating for Preserving Paper-Art-Works. J. Cult. Heritage 2011, 12, 380–383.

(16) Ogihara, H.; Xie, J.; Okagaki, J.; Saji, T. Simple Method for Preparing Superhydrophobic Paper: Spray-Deposited Hydrophobic Silica Nanoparticle Coatings Exhibit High Water-Repellency and Transparency. Langmuir 2012, 28, 4605–8.

(17) Wang, H.; Tang, L. M.; Wu, X. M.; Dai, W. T.; Qiu, Y. P. Fabrication and Anti-Frosting Performance of Super Hydrophobic Coating, Based on Modified Nano-sized Calcium Carbonate and Ordinary Polyacrylate. Appl. Surf. Sci. 2007, 253, 8818–8824.

(18) Samyn, P.; Schoukens, G.; Kiekens, P.; Mast, P.; Van den Abbeele, H.; Stanssens, D.; Vonck, L. Thermal Resistance of Organic Nanoparticle Coatings for Hydrophobicity and Water Repellence of Paper Substrates. Autex Res. J. 2010, 10, 100–109.

(19) Imani, R.; Talaiepour, M.; Dutta, J.; Ghobadinezhad, M. R.; Hemmasi, A. H.; Nazhad, M. M. Production of Antibacterial Filter Paper from Wood Cellulose. BioResources 2011, 6, 891–900.

(20) Martins, N. C. T.; Freire, C. S. R.; Pinto, R. J. B.; Fernandes, S. C. M.; Neto, C. P.; Silvestre, A. J. D.; Causio, J.; Baldi, G.; Sadocco, P.; Trindade, T. Electrostatic Assembly of Ag Nanoparticles onto Nanofibrillated Cellulose for Antibacterial Paper Products. Cellulose 2012, 19, 1425–1436.

(21) Ghule, K.; Ghule, A. V.; Chen, B. J.; Ling, Y. C. Preparation and Characterization of ZnO Nanoparticles Coated Paper and Its Antibacterial Activity Study. Green Chem. 2006, 8, 1034–1041.

(22) Dong, C. X.; Cairney, J.; Sun, Q. H.; Maddan, O. L.; He, G. H.; Deng, Y. L. Investigation of Mg(OH)2 Nanoparticles as an Antibacterial Agent. J. Nanopart. Res. 2010, 12, 2101–2109. Nanoparticles on Zinc Oxide Whiskers Incorporated in a Paper Matrix for Antibacterial Applications. J. Mater. Chem. 2009, 19, 2135–2140.

(24) Ngo, Y. H.; Li, D.; Simon, G. P.; Garnier, G. Paper Surfaces Functionalized by Nanoparticles. Adv. Colloid Interface Sci. 2011, 163, 23–38.

(25) Baruah, S.; Jaisai, M.; Imani, R.; Nazhad, M. M.; Dutta, J. Photocatalytic Paper Using Zinc Oxide Nanorods. Sci. Technol. Adv. Mater. 2010, 11, 1–7.

(26) Small, A. C.; Johnston, J. H. Novel Hybrid Materials of Cellulose Fibres and Doped ZnS Nanocrystals. Curr. Appl. Phys. 2008, 8, 512–515.

(27) Pelton, R.; Geng, X. L.; Brook, M. Photocatalytic Paper from Colloidal TiO2 Fact or Fantasy. Adv. Colloid Interface Sci. 2006, 127, 43–53.

(28) Munawar, R. F.; Zakaria, S.; Radiman, S.; Hua, C. C.; Abdullah, M.; Yamauchi, T. Properties of Magnetic Paper Prepared via in situ Synthesis Method. Sains Malays. 2010, 39, 593–598.

(29) Small, A. C.; Johnston, J. H. Novel Hybrid Materials of Magnetic Nanoparticles and Cellulose Fibers. J. Colloid Interface Sci. 2009, 331, 122–126.

(30) Anderson, R. E.; Guan, J. W.; Ricard, M.; Dubey, G.; Su, J.; Lopinski, G.; Dorris, G.; Bourne, O.; Simard, B. Multifunctional Single-Walled Carbon Nanotube-Cellulose Composite Paper. J. Mater. Chem. 2010, 20, 2400–2407.

(31) Agarwal, M.; Xing, Q.; Shim, B. S.; Kotov, N.; Varahramyan, K.; Lvov, Y. Conductive Paper from Lignocellulose Wood Microfibers Coated with a Nanocomposite of Carbon Nanotubes and Conductive Polymers. Nanotechnology 2009, 20, 1–8.

(32) Ihalainen, P.; Maattanen, A.; Jarnstrom, J.; Tobjork, D.; Osterbacka, R.; Peltonen, J. Influence of Surface Properties of Coated Papers on Printed Electronics. Ind. Eng. Chem. Res. 2012, 51, 6025–6036.

(33) Zhu, H.; Fang, Z.; Preston, C.; Li, Y.; Hu, L. Transparent Paper: Fabrications, Properties, and Device Applications. Energy Environ. Sci. 2014, 7, 269–287.

(34) Puurunen, K.; Vasara, P. Opportunities for Utilising Nanotechnology in Reaching Near-Zero Emissions in The paper Industry. J. Cleaner Prod. 2007, 15, 1287–1294.

(35) Shen, J.; Song, Z. Q.; Qian, X. R.; Yang, F.; Kong, F. G. Nanofillers for Papermaking Wet End Applications. BioResources 2010, 5, 1328–1331.

(36) Ibrahim, A. R.; Vuningoma, J. B.; Hu, X. H.; Gong, Y. N.; Hua, D.; Hong, Y. Z.; Wang, H. T.; Li, J. High-pressure Gas-Solid Carbonation Route Coupled with a Solid Ionic Liquid for Rapid Synthesis of Rhombohedral Calcite. J. Supercrit. Fluids 2012, 72, 78–83.

(37) Hu, Z. S.; Deng, Y. L.; Sun, Q. H. Synthesis of Precipitated Calcium Carbonate Nanoparticles using a Two-Membrane System. Colloid J. 2004, 66, 745–750.

(38) Chen, J. F.; Wang, Y. H.; Guo, F.; Wang, X. M.; Zheng, C. Synthesis of Nanoparticles with Novel Technology: High-Gravity Reactive Precipitation. Ind. Eng. Chem. Res. 2000, 39, 948–954.

(39) Carmona, J. G.; Morales, J. G.; Sainz, J. F.; Loste, E.; Clemente, R. R. The Mechanism of Precipitation of Chain-Like Calcite. J. Cryst. Growth 2004, 262, 479–489.

(40) Lei, M.; Li, P. G.; Sun, Z. B.; Tang, W. H. Effects of Organic Additives on the Morphology of Calcium Carbonate Particles in the Presence of CTAB. Mater. Lett. 2006, 60, 1261–1264.

(41) Li, S. X.; Yu, L.; Geng, F.; Shi, L. J.; Zheng, L. Q.; Yuan, S. L. Facile Preparation of Diversified patterns of Calcium Carbonate in the Presence of DTAB. J. Cryst. Growth 2010, 312, 1766–1773.

(42) Huang, J. H.; Mao, Z. F.; Luo, M. F. Effect of Anionic Surfactant on Vaterite CaCO3. Mater. Res. Bull. 2007, 42, 2184–2191.

(43) Mao, Z. F.; Huang, J. H. Habit Modification of Calcium Carbonate in the Presence of Malic Acid. J. Solid State Chem. 2007, 180, 453–460.

(44) Maeda, H.; Kasuga, T. Preparation of Poly(lactic acid) Composite Hollow Spheres Containing Calcium Carbonates. Acta Biomater. 2006, 2, 403–408.

(45) Wang, C. Y.; Piao, C.; Zhai, X. L.; Hickman, F. N.; Li, J. Synthesis and Characterization of Hydrophobic Calcium Carbonate Particles via a Dodecanoic Acid Inducing Process. Powder Technol. 2010, 198, 131–134.

(46) Sheng, Y.; Zhou, B.; Wang, C. Y.; Zhao, X.; Deng, Y. H.; Wang, Z. C. In Situ Preparation of Hydrophobic CaCO3 in the Presence of Sodium Oleate. Appl. Surf. Sci. 2006, 253, 1983–1987.

(47) Sheng, Y.; Zhou, B.; Zhao, J. Z.; Tao, N.; Yu, K. F.; Tian, Y. M.; Wang, Z. C. Influence of Octadecyl Dihydrogen Phosphate on the Formation of Active Super-Fine Calcium Carbonate. J. Colloid Interface Sci. 2004, 272, 326–329.

(48) Hoang, V. T.; Lam, D. T.; Hoang, D. V.; Hoang, T. Facile Surface Modification of Nanoprecipitated Calcium Carbonate by Adsorption of Sodium Stearate in Aqueous Solution. Colloids Surf., A 2010, 366, 95–103.

(49) Deepika; Hait, S. K.; Christopher, J.; Chen, Y.; Hodgson, P.; Tuli, D. K. Preparation and Evaluation of Hydrophobically Modified Core Shell Calcium Carbonate Structure by Different Capping Agents. Powder Technol. 2013, 235, 581–589.

(50) Chen, X.; Zhu, Y. C.; Guo, Y. P.; Zhou, B.; Zhao, X.; Du, Y. Y.; Lei, H.; Li, M. G.; Wang, Z. C. Carbonization Synthesis of Hydrophobic CaCO3 at Room Temperature. Colloids Surf., A 2010, 353, 97–103.

(51) Yu, J. G.; Lei, M.; Cheng, B.; Zhao, X. J. Effects of PAA Additive and Temperature on Morphology of Calcium Carbonate Particles. J. Solid State Chem. 2004, 177, 681–689.

(52) Szczes, A.; Chibowski, E.; Holysz, L. Influence of Ionic Surfactants on the Properties of Freshly Precipitated Calcium Carbonate. Colloids Surf., A 2007, 297, 14–18.

(53) Pal, M. K.; Gautam, J. Synthesis and Characterization of Polyacrylamide-Calcium Carbonate and Polyacrylamide-Calcium Sulfate Nanocomposites. Polym. Compos. 2012, 33, 515–523.

(54) Matahwa, H.; Ramiah, V.; Sanderson, R. D. Calcium Carbonate Crystallization in the Presence of Modified Polysaccharides and Linear Polymeric Additives. J. Cryst. Growth 2008, 310, 4561–4569.

(55) Kontrec, J.; Kralj, D.; Brecevic, L.; Falini, G. Influence of Some Polysaccharides on the Production of Calcium Carbonate Filler Particles. J. Cryst. Growth 2008, 310, 4554–4560.

(56) Yu, J. G.; Zhao, X. F.; Cheng, B.; Zhang, Q. J. Controlled Synthesis of Calcium Carbonate in a Mixed Aqueous Solution of PSMA and CTAB. J. Solid State Chem. 2005, 178, 861–867.

(57) Nan, Z.; Shi, Z.; Yan, B.; Guo, R.; Hou, W. A Novel Morphology of Aragonite and an Abnormal Polymorph Transformation from Calcite to Aragonite with PAM and CTAB as Additives. J. Colloid Interface Sci. 2008, 317, 77–82.

(58) Yan, G. W.; Wang, L.; Huang, J. H. The Crystallization Behavior of Calcium Carbonate in Ethanol/Water Solution Containing Mixed Nonionic/Anionic Surfactants. Powder Technol. 2009, 192, 58–64.

(59) Zhang, C. X.; Zhang, J. L.; Feng, X. Y.; Li, W.; Zhao, Y. J.; Han, B. X. Influence of Surfactants on the Morphologies of CaCO3 by Carbonation Route with Compressed CO2. Colloids Surf., A 2008, 324, 167–170.

(60) Pang, P.; Modgi, S. B.; Englezos, P. Optimisation of the Use of Calcium Carbonate as Filler in Mechanical Pulp. Appita J. 2003, 56, 122–126.

(61) Kim, C. H.; Cho, S. H.; Park, W. P. Inhibitory Effect of Functional Packaging Papers Containing Grapefruit Seed Extracts and Zeolite Against Micobial Growth. Appita J. 2005, 58, 202–207.

(62) Chuahan, V. S.; Bhardway, N. K.; Chakrabarti, S. K. Inorganic Filler – Modification and Retention During Papermaking: A Review. Appita J. 2011, 23, 93–100.

(63) Kim, D. S.; Lee, C. K. Surface Modification of Precipitated Calcium Carbonate Using Aqueous Fluosilicic Acid. Appl. Surf. Sci. 2002, 202, 15–23.

(64) Wang, J.; Chen, J. S.; Zong, J. Y.; Zhao, D.; Li, F.; Zhuo, R. X.; Cheng, S. X. Calcium Carbonate/Carboxymethyl Chitosan Hybrid Microspheres and Nanospheres for Drug Delivery. J. Phys Chem. C 2010, 114, 18940–18945.

(65) Zhao, Y. L.; Hu, Z. S.; Ragauskas, A.; Deng, Y. L. Improvement of Paper Properties Using Starch-Modified Precipitated Calcium

Carbonate Filler. Tappi J. 2005, 4, 3–7.

(66) Zhang, S. C.; Li, X. G. Synthesis and Characterization of CaCO3@SiO2 Core-Shell Nanoparticles. Powder Technol. 2004, 141, 75–79.

(67) Ukrainczyk, M.; Kontrec, J.; Kralj, D. Precipitation of Different Calcite Crystal Morphologies in the Presence of Sodium Stearate. J. Colloid Interface Sci. 2009, 329, 89–96.

(68) Sze, A.; Erickson, D.; Ren, L. Q.; Li, D. Q. Zeta-Potential Measurement Using the Smoluchowski Equation and the Slope of the Current−Time Relationship in Electroosmotic Flow. J. Colloid Interface Sci. 2003, 261, 402–410.

(69) Aloui, H.; Khwaldia, K.; Ben Slama, M.; Hamdi, M. Effect of Glycerol and Coating Weight on Functional Properties of Biopolymer-Coated Paper. Carbohydr. Polym. 2011, 86, 1063–1072.

(70) Borch, J. In Handbook of Physical Testing of Paper, 2nd ed.; Mark, R. E., Ed.; Marcel Dekker: New York, 2001; Part 3, pp 267–298.

(71) Bonham, J. S. The Appearance of 'White' Papers. Appita J. 2006, 59, 446–451.

(72) Wang, C. Y.; Sheng, Y.; Zhao, X.; Pan, Y.; Hari-Bala; Wang, Z. C. Synthesis of Hydrophobic CaCO3 Nanoparticles. Mater. Lett. 2006, 60, 854–857.

(73) Forbes, T.Z.; Radha, A.V.; Navrotsky, A. The Energetics of Nanophase Calcite. Geochim. Cosmochim. Acta 2011, 75, 7893–7905.

(74) Shentu, B. Q.; Li, J. P.; Weng, Z. X. Effect of Oleic Acid-Modified Nano-CaCO3 on the Crystallization Band Mechanical Properties of Polypropylene. Chin. J. Chem. Eng. 2006, 14, 814–818.

(75) Domingo, C.; Loste, E.; Gomez-Morales, J.; Garcia-Carmona, J.; Fraile, J. Calcite Precipitation by a High-Pressure CO2 Carbonation Route. J. Supercrit. Fluids 2006, 36, 202–215.

(76) Lohmander, S.; Rigdahl, M. Influence of a Shape Factor of Pigment Particles on the Rheological Properties of Coating Colours. Nord. Pulp Pap. Res. J. 2000, 15, 231–236.

(77) Lohmander, S. Influence of Shape and a Shape Factor of Pigment Particles on the Packing Ability in Coating Layers. Nord. Pulp Pap. Res. J. 2000, 15, 300–305.

(78) Nypelo, T.; Osterberg, M.; Zu, X. J.; Laine, J. Preparation of Ultrathin Coating Layers Using Surface Modified Silica Nanoparticles. Colloids Surf., A 2011, 392, 313–321.